# On Preparing a List of Random Treatment Assignments: A New Computer-Aided Design

N. S. Santos-Magalhaes, H. M. de Oliveira, A. J. Alves[*]

School of Pharmacy, Federal University of Pernambuco, Brazil

*Abstract - This paper presents the foundations of a computer oriented approach for preparing a list of random treatment assignments to be adopted in randomised controlled trials. Software is presented which can be applied in the earliest stage of clinical trials and bioequivalence assays. This allocation of patients to treatment in clinical trials ensures exactly equal treatment numbers. The investigation of the randomness properties of an assignment leads to the concept of a "strong randomised list". The new approach introduced in this note is based on thresholds and produces a strong randomised list of treatment assignments.*

## 1. INTRODUCTION

The Pharmaceutical industry has had a huge interest in organising trials to resolve relevant therapeutic aspects. Clinical trials play a very important role in the evaluation of new issues for drug therapy. Statistics is not only decisive in data analysis but also beforehand the study's design of a trial. It has long been recognised that uncontrolled trial potentially provide distorted outcome of the therapy. In contrast, clinical trials with a properly randomised control groups avoid bias and provide a basis for statistical tests. One essential is that clinician must be unable to predict what the assignment will be. Simple randomisation, replacement randomisation, biased coin method and random permuted blocks are some approaches normally adopted to generate a list of assignments[1-5]. This paper presents a computer-oriented approach for preparing a random list of treatment assignments. Statistical properties of randomisation lists are also examined.

## 2. THE RANDOMIZATION LIST GENERATION

The use of a software to produce lists based on a computer generated sequence of pseudo-random numbers[6] is a fully convenient tool for staffs extensively involved in clinical trials. Properly randomised control groups avoid drawbacks of the simple randomisation and the replacement randomisation[1-4]. It is therefore helpful to design randomisation to ensure similar treatment numbers throughout a trial. The computer generation of a random assignment introduced in this paper is based on an internal (0,1)-uniform distributed random number generator[6]. The list can be used one at a time as patients are registered into the trial. Some care should be taken to avoid repeating the same output list by assuring different lists of assignments in each program run. A clock-dependent seed must therefore be adopted. Let N be the number of patients registered into the trial and let g denote the number of patient groups. U denotes the uniformed-distributed random variable. The procedure is flexible allowing an experiment design for any number of patients and any number of groups with an equal number of patients provided that g divides N. The approach is founded on a choice of a threshold set $\{T_1,T_2,T_3...,T_N\}$. The key point is that the random sampling is made without reposition. Two lists (vectors) are then considered:

i) A list $\vec{L}$ of assignments $\vec{L}=\{L_1,L_2,L_3,...,L_N\}$ and

ii) A list $\vec{S}$ of survivors $\vec{S}=\{S_1,S_2,S_3,...,S_N\}$.

___________________________________

[*] Correspondence to: Prof. N.S. Santos Magalhães - *Laboratório de Tecnologia Químico-Farmacêutica* LTQF, Departamento de Farmácia, Universidade Federal de Pernambuco, UFPE, Av. Prof. Arthur Sá, s/n, 50.740-520 Recife - PE Brazil. Fax. 55 81 271-8559, E-mail: nssm@ufpe.br,  hmo@ufpe.br and  leac@nlink.com.br

The procedure begins with an empty list of assignments, $\vec{L}=\{\ \}$, and a full list of survivors, $\vec{S}=\{1,2,3,...,N\}$. Pick up then the first random number $U_1$. Setting the thresholds as $T_k=k/N$, a patient number k is chosen *if and only if* $T_{k-1} \leq U_1 < T_k$. Therefore k is included into the list $\vec{L}$, i.e., $L_1 \leftarrow k$ (the mean of L←k being the usual in computer languages: assign the value k to the variable L). Thus, the probability of selecting a *particular* patient number k is uniformed distributed, $P(k)=1/N$ ($\forall k$). Since the random sample must be taken with no reposition, the patient number k must be deleted in the survivor list $\vec{S}$. This circumvents considering the index k in the next step. The likelihood of selecting an arbitrary patient number (except k) must be updated using conditional probabilities since that the sample space had changed[7]. The thresholds are adjusted to be now $T_k=k/(N-1)$, k=1,2,...,N-1. A new choice of the list is done by a second random guess $U_2$, proceeding a comparison with the up-dated set of thresholds. The procedure goes on in a similar way in each new step and the thresholds for the I[th]-step are $T_k=k/\|\vec{S}\|$. The probability of selecting a *particular* patient number k is uniformed distributed and the conditional probability $P(k|\vec{S})=P(T_{k-1} \leq U_I < T_k | \vec{S})= 1/\|\vec{S}\|$ so that the following algorithm is established:

## 3. THE THRESHOLD-BASED ALGORITHM

BEGIN
- P1. Read N and g. If g | N continue else enter data again (g | N denotes g divides N).
- P2. Set initial conditions: I=0, $L_i$=0, $S_i$=i ($\forall$i=1,N), i.e. $\vec{L}$={0,0,0,...,0} and $\vec{S}$={1,2,3,...,N}.
- P3. WHILE I<N DO
  BEGIN
  - P3.1. Threshold up dating: $T_k = \dfrac{k}{N-I}$, k=0,1,2,...,N-I.
    (There is exactly N-I thresholds)
  - P3.2. Next step: I $\leftarrow$ I+1.
  - P3.3. Pick up the I$^{th}$-uniform random number U.
  - P3.4. If $T_{k-1} \leq U_I < T_k$ (k=1,2,...,N-I) the k must be included in the random list $\vec{L}$, that is, $L_I \leftarrow$ k.
  - P3.5. Up-date (shrink) the survivor list.
    (*Since the random list should have no repetition, k must be deleted from the survivor list.*)
  END;
- P4. FOR m equal 1 to g:
  Print the group number and their corresponding elements;

END.

After the random generation of the list $\vec{L}$, it is divided into g groups namely

$G_1 = \{L_1, L_2, ..., L_{N/g}\}$, $G_2 = \{L_{N/g+1}, L_{N/g+2}, ..., L_{2N/g}\}$, ..., $G_g = \{L_{N-N/g+1}, ..., L_{N-1}, L_N\}$

each of them with cardinality $\|G_i\|$=N/g  i=1,2,3...,g.

A naive illustrative example is presented in the sequel showing how to design a simple randomised trial with 12 patients and 2 treatments.

| Random number | list of assignments $\vec{L}$ | survivor list $\vec{S}$ |
|---|---|---|
| $U_1$=0.168502561... | L={3} | S={1,2,4,5,6,7,8,9,10,11,12} |
| $U_2$=0.658033330... | L={3,9} | S={1,2,4,5,6,7,8,10,11,12} |
| $U_3$=0.093729293... | L={3,9,1} | S={2,4,5,6,7,8,10,11,12} |
| $U_4$=0.756609143... | L={3,9,1,10} | S={2,4,5,6,7,8,11,12} |
| $U_5$=0.463955829... | L={3,9,1,10,6} | S={2,4,5,7,8,11,12} |
| $U_6$=0.070162761... | L={3,9,1,10,6,5} | S={2,4,7,8,11,12} |
| $U_7$=0.222246588... | L={3,9,1,10,6,5,11} | S={2,4,7,8,12} |
| $U_8$=0.706319757... | L={3,9,1,10,6,5,11,8} | S={2,4,7,12} |
| $U_9$=0.586996776... | L={3,9,1,10,6,5,11,8,12} | S={2,4,7} |
| $U_{10}$=0.752142819... | L={3,9,1,10,6,5,11,8,12,7} | S={2,4} |
| $U_{11}$=0.937703174... | L={3,9,1,10,6,5,11,8,12,7,4} | S={2} |
| $U_{12}$=0.669782608... | L={3,9,1,10,6,5,11,8,12,7,4,2} | S={ } |

Threshold calculations:

| T[1] | T[2] | T[3] | T[4] | T[5] | T[6] | T[7] | T[8] | T[9] | T[10] | T[11] | T[12] |
|------|------|------|------|------|------|------|------|------|-------|-------|-------|
| 0.08$\underline{33}$ | 0.01$\underline{66}$ | 0.25 | 0.$\underline{33}$ | 0.41$\underline{66}$ | 0.5 | 0.58$\underline{33}$ | 0.$\underline{66}$ | 0.75 | 0.8$\underline{33}$ | 0.91$\underline{66}$ | 1.00 |
| 0.0$\underline{90}$ | 0.$\underline{18}$ | 0.$\underline{27}$ | 0.$\underline{36}$ | 0.$\underline{45}$ | 0.$\underline{54}$ | 0.$\underline{63}$ | 0.$\underline{72}$ | 0.$\underline{81}$ | 0.$\underline{90}$ | 1.00 | |
| 0.1 | 0.2 | 0.3 | 0.4 | 0.5 | 0.6 | 0.7 | 0.8 | 0.9 | 1.00 | | |
| 0.$\underline{11}$ | 0.$\underline{22}$ | 0.$\underline{33}$ | 0.$\underline{44}$ | 0.$\underline{55}$ | 0.$\underline{66}$ | 0.$\underline{77}$ | 0.$\underline{88}$ | 1.00 | | | |
| 0.125 | 0.25 | 0.375 | 0.5 | 0.625 | 0.75 | 0.875 | 1.00 | | | | |

...
...

0.5  1.00
1.00

The $I^{th}$ set of threshold is given by T[k]=k/(12-I), k=1,2,...,12-I and the underline denotes an infinite repetition, e.g. 0.08$\underline{33}$=0.08333333333...

For instance, the 4$^{th}$ patient assignment depends only on $U_4$=0.756609143... The survivor list prior this step is $\vec{S}$ ={2,4,5,6,7,8,10,11,12}. Since that T[6]=0.66 ≤ $U_4$<T[7]=0.77 the 7$^{th}$ element of the survivor list is selected: Insert number 10 in the list of assignments and up-date the survivor list by erasing such a patient number: $\vec{S}$={2,4,5,6,7,8,11,12}.

The treatment (A or B) is decided after obtaining the complete list of assignments:

$$\vec{L} = \{3,9,1,10,6,5,11,8,12,7,4,2\}.$$

Treatment A: The first N/g=6 patients;

Treatment B: The last N/g=6 patients of the list $\vec{L}$.

## 4. EXAMINING THE RANDOMNESS OF A LIST OF ASSIGNMENTS

The following propositions corroborate that the threshold-based approach achieves properly randomised groups. The symbols ∪ and ∩ denote the union and intersection of events, respectively.

**Proposition 1**. An arbitrary patient (say number k) has exactly the same probability to be in any position of the list of assignments, that is,

$$P(L_i = k) = \frac{1}{N} \quad (\forall k = 1,2,...,N), \quad (\forall i = 1,2,...N). \quad \blacksquare$$

Proof. Since that U is a uniform random variable, $P(L_1=k)=P(0<U_1<1/N)=\frac{1}{N}$. It is worthwhile to remark then that $P(L_2 = k) = P(L_2 = k \cap L_1 \neq k)$ so that

$$P(L_2 = k) = P(L_1 \neq k) P(L_2 = k | L_1 \neq k) = \frac{N-1}{N} \frac{1}{N-1} = \frac{1}{N}.$$

Following, the probability of the patient number k be assigned in the 3$^{rd}$ position of the list is evaluated: $P(L_3 = k) = P(L_3 = k \cap L_1 \neq k \cap L_2 \neq k)$. Therefore,

$P(L_3 = k) = P(L_1 \neq k)P(L_2 \neq k | L_1 \neq k)P(L_3 = k | L_1 \neq k, L_2 \neq k) =$

$\frac{N-1}{N} \frac{N-2}{N-1} \frac{1}{N-2} = \frac{1}{N}$. By using a similar reasoning, it can be proved that

$P(L_4 = k) = \frac{N-1}{N} \frac{N-2}{N-1} \frac{N-3}{N-2} \frac{1}{N-3} = \frac{1}{N}$ and the proof is completed by induction

QED.

**Proposition 2**. An arbitrary patient has exactly the same probability to be in a treatment, no matter which the treatment is. ∎

Proof. Let us consider an arbitrary treatment, say m. The set $G_m$ is the list of patients under the treatment m, $1 \leq m \leq g$. The probability of the $k^{th}$ patient be assigned to such a treatment is given by

$P(k \in G_m) = P(\bigcup_{i=(m-1)\frac{N}{g}+1}^{m \cdot \frac{N}{g}} \{L_i = k\})$. Since the events are mutually exclusive, it follows

that $P(k \in G_m) = \sum_{i=(m-1)\frac{N}{g}+1}^{m \cdot \frac{N}{g}} P(L_i = k)$. Using proposition 1 completes the proof:

$P(k \in G_m) = \sum_{i=(m-1)\frac{N}{g}+1}^{m \cdot \frac{N}{g}} \frac{1}{N} = \|G_m\| \frac{1}{N} = \frac{N}{g} \frac{1}{N} = \frac{1}{g}$. QED.

For a number of g treatments, a number of g random groups are firstly generated. The clinical trial involves g phases designed in such a way that each group is submitted to all the g treatments, one at a time. The randomisation list of consecutive random treatments must now be generated. In other words, after defining the patients groups by the threshold approach, it is required to decide which treatment must be administrated to each group, on each study phase (phase 1,2, ..., g). The threshold method is called once more so as to find the phase-I (random) assignment. Simply considering all cyclic permutations[8] of the first assignment for phase-I can complete the design of the next phases. A g × g matrix [G] is defined in which the phase-I assignment fulfil the first row. The other rows are cyclic permutations of the first row, so the elements of such matrix can be computed in terms of the first row elements according to:

G(i,j)=G(1, { ϕᵢ(j) } MOD g), i=2,3,...,g and j=1,2,3,...,g

Where $\phi_i(j) = \begin{cases} g & j = i-1 \\ j-(i-1) \text{ MOD } g, & \text{otherwise.} \end{cases}$

The operation "MOD g" denotes the classical modulo g reduction[8].

The final list of the random trial assignment is given by the matrix:

$$[G] = \begin{pmatrix} G(1,1) & G(1,2) & \cdots & G(1,g) \\ G(2,1) & G(2,2) & \cdots & G(2,g) \\ \vdots & \vdots & \ddots & \vdots \\ G(g,1) & G(g,2) & \cdots & G(g,g) \end{pmatrix}$$

On the phase i, the group G(i,j) is submitted to the treatment number j. As an example, a 4-treatment case is considered in the sequel. If the four uniform distributed random variables at the output of the threshold assignment furnishes $U_1=0.6531...U_2=0.1497...U_3=0.7121... U_4=0.2437...$, then the assignment list will be:

$1/2 \leq U_1 < 3/4$    $\vec{L} = \{3\}$
$0 \leq U_2 < 1/3$    $\vec{L} = \{3,1\}$
$1/2 \leq U_3 < 1$    $\vec{L} = \{3,1,4\}$
$0 \leq U_4 < 1$    $\vec{L} = \{3,1,4,2\}$.

Therefore, the phase-I assignment is 3 1 4 2, which means: Group #3: treatment 1; Group #1: treatment 2; Group #4: treatment 3; Group #2: treatment 4. All the cyclic permutations of the phase-I assignment are now generated yielding a 4×4 matrix:

treatment

$$\begin{pmatrix} 3 & 1 & 4 & 2 \\ 2 & 3 & 1 & 4 \\ 4 & 2 & 3 & 1 \\ 1 & 4 & 2 & 3 \end{pmatrix} \text{ phase}$$

Thus, on the phase II, the group 2 is submitted to treatment 1, the group 3 is submitted to treatment 2 and so on.

**Proposition 3.** The probability of an arbitrary treatment be assigned to any group at any phase is exactly the same. ∎

Proof. The probability of the $m^{th}$ group be assigned to the $j^{th}$ treatment on the phase i is P(G(i,j)=m). On the other hand, since that the phase I assignment was derived from the threshold approach, it follows that

$$P(G(1,j) = m) = \frac{1}{g} \quad (\forall j).$$

Remembering now that cyclic permutations were used to derive the phase-i

assignment, i>1, then:

$$P(G(i,j) = m) = P(G(1, \{\phi_i(j)\} \text{ MOD } g) = m).$$

Therefore, given an arbitrary group $G_m$, the probability of $G_m$ being selected for any treatment j is the same, no matter the phase, i.e.

$$P(G(i,j) = m) = \frac{1}{g} \quad (\forall i) \ (\forall j) \ (\forall m). \text{ QED.}$$

The randomisation list is finally transferred to a sequence of sealed envelopes. Indeed it is supposed that no information at all on groups' order should be furnished, otherwise such knowledge could be used and the conditional probabilities will be:

$$P(G(i,j) = m \mid G(1,1),...G(1,g)) = P(G(1,\phi_i(j)) = m \mid G(1,1),...G(1,g)) = \begin{cases} 1 & \text{if } m = G(1, \phi_i(j)), \\ 0 & \text{otherwise.} \end{cases}$$

for i≥2, that is, the next phase assignments will be deterministic.

Finally, the following concept in the random lists design is introduced:

**Definition**. A randomisation list of consecutive random treatments prepared in such a way that Propositions 1 to 3 hold is defined as a *strong randomisation assignment*. ∎

## 5. DISCUSSION

Clinical trials are one of the most adopted strategies in the evaluation of new issues for drug therapy. A computer-aided approach for preparing a list of random treatment assignments is offered as a tool for groups extensively involved in such trials. This way of patients' selection ensures *exactly* equal treatment numbers. The technique, referred as to the "threshold approach", is general and can be applied for a randomised trial with *any* number of treatments. It can also easily be extended to stratified randomisation. Aiming to avoid that clinician "breaks the code" no information on the groups' order must be furnished in the final list. The randomness of the patient's allocation to treatment is examined showing that a "strong randomisation" is achieved. Freeware software for preparing a randomisation list is available on request at hmo@ufpe.br.

ACKNOWLEDGEMENTS

*This research had a partial financial support by the Pharmaceutical Laboratory of the State of Pernambuco - LAFEPE, Brazil. The first author also thanks Brazilian National Council for Scientific and Technological Development (CNPq) under grant #304946/85-1.*